\documentclass[10pt,twocolumn,aps,pra,amsmath,amssymb,showpacs]{revtex4-1}
\usepackage{bm}
\usepackage{mathrsfs}
\usepackage{graphicx}
\usepackage[usenames,dvipsnames,svgnames,table]{xcolor}
\usepackage[unicode=true,
            pdfusetitle, 
            bookmarks=true,
            bookmarksnumbered=false,
            bookmarksopen=false,
            breaklinks=true,
            pdfborder={0 0 0},
            backref=false,
            colorlinks=true]{hyperref}
\hypersetup{linkcolor=NavyBlue,urlcolor=NavyBlue,citecolor=NavyBlue}

\usepackage{amsthm}

\providecommand{\propositionname}{Proposition}

\usepackage{tikz-cd}

\newcommand{\cH}{\mathcal{H}}

\begin{document}

\title{Reference state for arbitrary $U$-consistent subspace}
\author{Iman Sargolzahi}
\email{sargolzahi@neyshabur.ac.ir; sargolzahi@gmail.com}
\affiliation{Department of Physics, University of Neyshabur, Neyshabur, Iran}
%\author{Sayyed Yahya Mirafzali}
%\email{y.mirafzali@vru.ac.ir}
%\affiliation{Department of Physics, Faculty of Science, Vali-e-Asr University of Rafsanjan, Rafsanjan, Iran}

\begin{abstract}
The reduced dynamics of the system $S$, interacting with the environment $E$, is not given by a linear map, in general. However, if it is given by a linear map, then this map is also Hermitian. In order that the reduced dynamics of the system is given by a linear Hermitian map, there must be some restrictions on the set of possible initial states of the system-environment or on the possible unitary evolutions of  the whole $SE$.
In this paper, adding an ancillary reference space $R$, we assign to each convex  set of possible initial states of  the system-environment $\mathcal{S}$, for which the reduced dynamics is Hermitian, a tripartite state  $\omega_{RSE}$, which we call it the \textit{reference state}, such that 
 the set $\mathcal{S}$ is given as the steered states from the  reference state $\omega_{RSE}$,.  The set of possible initial states of the system is also given as the steered set from
a bipartite reference state $\omega_{RS}$. The relation between these two reference states is as $\omega_{RSE}=id_{R}\otimes \Lambda_{S}(\omega_{RS})$, where $id_{R}$ is the identity map on $R$ and $\Lambda_{S}$ is a Hermitian assignment map, from $S$ to $SE$.
%Finally, during the study of an example, we show that for non-completely positive-ness of the reduced dynamics, not only the non-completely positive-ness of the assignment map is required, but also the non-Markovianity of 
% the reference state $\omega_{RSE}$, is needed.
As an important consequence of introducing the reference state  $\omega_{RSE}$, we generalize the result of 
 [F. Buscemi, \href{http://dx.doi.org/10.1103/PhysRevLett.113.140502} {Phys. Rev. Lett. {\bf 113}, 140502 (2014)}]: We show that, for a \textit{$U$-consistent} subspace, the reduced dynamics of the system is completely positive, for arbitrary unitary evolution of the whole system-environment $U$, if and only if the reference state $\omega_{RSE}$ is a Markov state.
 In addition, we show that the evolution of the set of system-environment (system) states is determined by the evolution of the reference state $\omega_{RSE}$ ($\omega_{RS}$).

\end{abstract}

%\pacs{03.65.Yz, 03.67.-a}

\maketitle
%Introduction
\section{Introduction} \label{sec:intro}

Consider a closed finite dimensional quantum system which evolves as
\begin{equation}
\label{eq:one}
\begin{aligned}
\rho\rightarrow\rho^{\prime} =Ad_{U}(\rho)\equiv U\rho U^{\dagger},
\end{aligned}
\end{equation}
where $\rho$ and $\rho^{\prime}$ are the initial and final states (density operators) of the system, respectively, and $U$ is a unitary operator ($UU^{\dagger}= U^{\dagger}U=I$, where $I$ is the identity operator).

In general, the system is not closed and interacts with its environment. We can consider the whole system-environment as a closed quantum system which evolves as Eq. (\ref{eq:one}). So the reduced state of the system after the evolution is given by
\begin{equation}
\label{eq:two}
\begin{aligned}
\rho_{S}^{\prime}=\mathrm{Tr_{E}} \circ Ad_{U}(\rho_{SE})=\mathrm{Tr_{E}}\left( U \rho_{SE}U^{\dagger}\right), 
\end{aligned}
\end{equation} 
where $\rho_{SE}$ is the initial state of the combined system-environment quantum system and $U$ acts on the whole Hilbert space of the system-environment. 

In general, the relation between the initial state of the system $\rho_{S}=\mathrm{Tr_{E}}(\rho_{SE})$ and its final state $\rho_{S}^{\prime}$ is not given by a map \cite{2, 3}. Even if it is given by a map, then, in general, this map is not a linear map \cite{4, 5}.
In order that Eq. (\ref{eq:two}) leads to a linear map from $\rho_{S}$ to $\rho_{S}^{\prime}$, there must be some restrictions on the set of possible initial states of the system-environment  $\lbrace\rho_{SE}\rbrace$ or on the possible unitary evolutions  $U$ \cite{3, 6}.

However, if the reduced dynamics of the system from $\rho_{S}$ to $\rho_{S}^{\prime}$ can be given by a linear map $\Psi$, then  this $\Psi$ is also Hermitian, i.e., maps each Hermitian operator to a Hermitian operator.
Now, an important result is that for each linear trace-preserving Hermitian map from $\rho_{S}$ to $\rho_{S}^{\prime}$, there exists an operator sum representation  in the following form:
\begin{equation}
\label{eq:three}
\begin{aligned}
\rho_{S}^{\prime}=\sum_{i}e_{i}\,\tilde{E_{i}}\,\rho_{S}\,\tilde{E_{i}}^{\dagger},\ \ \ \sum_{i}e_{i}\,\tilde{E_{i}}^{\dagger}\tilde{E_{i}}=I_{S},
\end{aligned}
\end{equation}
where $\tilde{E_{i}}$ are linear operators and $e_{i}$ are real coefficients \cite{8, 7, 3}.
For the special case that all of the coefficients $e_{i}$ in Eq. (\ref{eq:three}) are positive, then we call the map completely positive (CP) \cite{1} and rewrite Eq. (\ref{eq:three}) in the following form:  
\begin{equation}
\label{eq:four}
\begin{aligned}
\rho_{S}^{\prime}=\sum_{i}E_{i}\,\rho_{S}\,E_{i}^{\dagger},\ \ \ \sum_{i}E_{i}^{\dagger}E_{i}=I_{S},
\end{aligned}
\end{equation}
where $E_{i}\equiv\sqrt{e_{i}}\,\tilde{E_{i}}$.

A general framework for linear trace-preserving Hermitian maps, arisen from  Eq. (\ref{eq:two}), when both the system and the environment are finite dimensional, has been developed in Ref. \cite{3}. The starting point of this framework is to consider a \textit{convex} set of initial states $\mathcal{S}=\lbrace \rho_{SE} \rbrace$, for the whole system-environment, i.e., if $\rho_{SE}^{(1)} , \rho_{SE}^{(2)} \in \mathcal{S}$, then $  \rho_{SE}=p\rho_{SE}^{(1)}+(1-p)\rho_{SE}^{(2)} \in \mathcal{S}$, for $0\leq p\leq 1$.

As we will see in the next section, an straightforward way to construct a convex $\mathcal{S}$ is to consider the set of \textit{steered} states from performing  measurements on the part $R$ of a fixed tripartite state $\omega_{RSE}$, which is a state on the Hilbert space of the reference-system-environment $\cH_{R}\otimes\cH_{S}\otimes\cH_{E}$. 

We call  $\omega_{RSE}$ the \textit{reference state} and we will show that if it can be written as Eq.  (\ref{eq:2a}) below, then the reduced dynamics of the system is Hermitian. 
Interestingly, this result includes all the previously found sets  $\mathcal{S}$, in Refs. \cite{11, 12, 14, 13, 15, 16}, for which the reduced dynamics of the system is CP.

Then, we question whether it is possible to find such reference state $\omega_{RSE}$ for arbitrary convex set $\mathcal{S}$, for which the reduced dynamics is Hermitian. Fortunately, this is the case as we will show in Sec. ~\ref{sec:ref state}. The possibility of introducing the reference state $\omega_{RSE}$, for arbitrary  $\mathcal{S}$,  has an important consequence: In Sec. ~\ref{sec:CP}, we generalize the result of Ref. \cite{15}, i.e., we show that, for arbitrary  $\mathcal{S}$, when $\omega_{RSE}$ is not a so-called  \textit{Markov state}, then the the reduced dynamics of the system, for at least one $U$, is not CP.

 Sections  ~\ref{sec:ref state} and ~\ref{sec:CP} are on the case that there is a one to one  correspondence between the members of  $\mathcal{S}$ and the members of $\mathrm{Tr_{E}}\mathcal{S}$. The general case, where there is no such correspondence, is given in Sec. ~\ref{sec:generalization}.
 
 In Sec. ~\ref{sec:example}, we consider the case studied in Ref. \cite{5}, as an example, to illustrate (a part of) our results, and 
%  and show that, as  the non-completely positive-ness of the assignment map, the non-Markovianity of 
% the reference state $\omega_{RSE}$, is also a necessary (but, in general, insufficient) condition for 
%the non-completely positive-ness of the reduced dynamics of the system $S$.
finally,
we will end this paper in  Sec.~\ref{sec:summary}, with a summary of our results.

\section{Reduced dynamics for a steered set} \label{sec:steered set}
%Reduced dynamics for a steered set

Assume that, for each $\rho_{SE} \in \mathcal{S}$, the reduced dynamics of the system is given by
a  map $\Psi$.  So, for each $\rho_{S}\in\mathrm{Tr}_{E}\mathcal{S}$, we have: 
\begin{equation}
\label{eq:eleven}
\begin{aligned}
\rho_{S}^{\prime}=\Psi(\rho_{S})=\mathrm{Tr_{E}} \circ Ad_{U}(\rho_{SE}). \\   
\end{aligned}
\end{equation}
The first obvious requirement that such a map $\Psi$ can be defined, is the \textit{$U$-consistency} of the $\mathcal{S}$ \cite{3}, i.e., if for two states $\rho_{SE}^{(1)}, \, \rho_{SE}^{(2)}\in\mathcal{S}$, we have $\mathrm{Tr}_{E}(\rho_{SE}^{(1)})=\mathrm{Tr}_{E}(\rho_{SE}^{(2)})=\rho_{S}$, then we must have $\mathrm{Tr}_{E}\circ Ad_{U}(\rho_{SE}^{(1)})=\mathrm{Tr}_{E}\circ Ad_{U}(\rho_{SE}^{(2)})=\Psi(\rho_{S})$.

Interestingly, if $\mathcal{S}$ is convex and $U$-consistent, then the reduced dynamics of the system is given by a (linear trace-preserving) Hermitian map \cite{3}.

An straightforward way to construct a convex $\mathcal{S}$ is to consider the set of steered states from performing  measurements on the part $R$ of a reference state $\omega_{RSE}$
%, which is a state on the Hilbert space of the reference-system-environment $\cH_{R}\otimes\cH_{S}\otimes\cH_{E}$
 \cite{ 15, 15a}:
\begin{equation}
\label{eq:1a}
\begin{aligned}
\mathcal{S}=\left\lbrace \frac{\mathrm{Tr_{R}}[(P_{R}\otimes I_{SE})\omega_{RSE}]}{\mathrm{Tr}[(P_{R}\otimes I_{SE})\omega_{RSE}]} ,  P_{R}>0 \right\rbrace ,  
\end{aligned}
\end{equation}
where $P_{R}$ is  arbitrary positive operator on $\cH_{R}$ such that $\mathrm{Tr}[(P_{R}\otimes I_{SE})\omega_{RSE}]>0$ and $I_{SE}$ is the identity operator on $\cH_{S}\otimes\cH_{E}$. Note that, up to a positive  factor, $P_{R}$ can be considered as an element of a POVM. 
%We will call  $\omega_{RSE}$ as the \textit{reference state} of the set of system-environment states $\mathcal{S}$, in Eq.  (\ref{eq:1a}).

It can be shown simply that the set of initial states of the system-environment  $\mathcal{S}$, in Eq.  (\ref{eq:1a}), is convex. So, if, in addition, it be a $U$-consistent set, then the reduced dynamics of the system, for all $\rho_{SE}\in \mathcal{S}$ ($\rho_{S}\in \mathrm{Tr_{E}}\mathcal{S}$), is given by a Hermitian map, as Eq. (\ref{eq:three}).

Note that if there is a one to one correspondence between the members of  $\mathcal{S}$ and the members of $\mathrm{Tr_{E}}\mathcal{S}$, then, trivially, the $U$-consistency condition is satisfied. In other words,
if, for each $\rho_{S}\in \mathrm{Tr_{E}}\mathcal{S}$, there is only one $\rho_{SE}\in \mathcal{S}$ such that $\rho_{S}= \mathrm{Tr_{E}}(\rho_{SE})$, then the set $\mathcal{S}$ is $U$-consistent, for any arbitrary unitary evolution of the whole system-environment $U$.

Now, let's consider the case that the reference state  $\omega_{RSE}$ can be written as
\begin{equation}
\label{eq:2a}
\begin{aligned}
\omega_{RSE}=id_{R}\otimes \Lambda_{S} (\omega_{RS}),  
\end{aligned}
\end{equation}
where $\omega_{RS}=\mathrm{Tr_{E}}(\omega_{RSE})$, $id_{R}$ is the identity map on $\mathcal{L}(\cH_{R})$ and $\Lambda_{S}\,:\mathcal{L}(\cH_{S})\rightarrow\mathcal{L}(\cH_{S}\otimes\cH_{E})$ is a Hermitian map. ($\mathcal{L}(\cH)$ is the space of linear operators on $\cH$.) So, each $\rho_{SE}^{(i)}\in \mathcal{S}$, in Eq. (\ref{eq:1a}), can be written as
\begin{equation}
\label{eq:3a}
\begin{aligned}
\rho_{SE}^{(i)}=(\mathrm{Tr_{R}}\circ \mathcal{P}_{R}^{(i)})\otimes \Lambda_{S} (\omega_{RS}) \qquad  \\
= (\mathrm{Tr_{R}}\circ \mathcal{P}_{R}^{(i)})\otimes id_{SE} (\omega_{RSE}), 
\end{aligned}
\end{equation}
where, the map $\mathcal{P}_{R}^{(i)}$ on $\mathcal{L}(\cH_{R})$ is defined as $\mathcal{P}_{R}^{(i)}(A_{R})=P_{R}^{(i)}A_{R}$, for each $A_{R}\in \mathcal{L}(\cH_{R})$. In addition,
  without loss of generality, we have considered only those $P_{R}^{(i)}$, in  Eq. (\ref{eq:1a}), for which we have $\mathrm{Tr}[(P_{R}^{(i)}\otimes I_{SE})\omega_{RSE}]=1$. Therefore
\begin{equation}
\label{eq:4a}
\begin{aligned}
\rho_{S}^{(i)}=\mathrm{Tr_{E}}(\rho_{SE}^{(i)}) \qquad\qquad\qquad\qquad\qquad\qquad\qquad\quad \\
=(\mathrm{Tr_{R}}\circ \mathcal{P}_{R}^{(i)})\otimes (\mathrm{Tr_{E}}\circ\Lambda_{S}) (\omega_{RS})\qquad\quad\qquad \\
=  [(\mathrm{Tr_{R}}\circ \mathcal{P}_{R}^{(i)})\otimes id_{S}][id_{R}\otimes (\mathrm{Tr_{E}}\circ\Lambda_{S})](\omega_{RS}) \\
=(\mathrm{Tr_{R}}\circ \mathcal{P}_{R}^{(i)})\otimes id_{S}(\omega_{RS}),\qquad\qquad\qquad\qquad
\end{aligned}
\end{equation}
where, in the fourth line, we have used this fact that, according to Eq. (\ref{eq:2a}), we have $[id_{R}\otimes (\mathrm{Tr_{E}}\circ\Lambda_{S})](\omega_{RS})= \mathrm{Tr_{E}}[id_{R}\otimes\Lambda_{S}(\omega_{RS})]=\mathrm{Tr_{E}}(\omega_{RSE})=\omega_{RS}$.

Next, assume that $\lbrace S_{j}\rbrace$ is an orthonormal basis (according to the Hilbert-Schmidt inner product \cite{1}) for $\mathcal{L}(\cH_{S})$. So, we can decompose $\omega_{RS}$ as
 \begin{equation}
\label{eq:5a}
\begin{aligned}
\omega_{RS}=\sum_{j} R_{j}\otimes S_{j}, 
\end{aligned}
\end{equation}
where $ R_{j}$ are linear operators in $\mathcal{L}(\cH_{R})$. Therefore, from Eq. (\ref{eq:4a}), we have
 \begin{equation}
\label{eq:6a}
\begin{aligned}
\rho_{S}^{(i)}=\sum_{j} \mathrm{Tr} (P_{R}^{(i)}R_{j}) S_{j}=\sum_{j} a_{ij} S_{j},
\end{aligned}
\end{equation}
where $a_{ij}=\mathrm{Tr} (P_{R}^{(i)}R_{j})$. From Eqs. (\ref{eq:2a}) and (\ref{eq:5a}), we have
 \begin{equation}
\label{eq:7}
\begin{aligned}
\omega_{RSE}=\sum_{j} R_{j}\otimes \Lambda_{S} (S_{j}).
\end{aligned}
\end{equation}
So, from From Eqs. (\ref{eq:3a}) and (\ref{eq:6a}), we get
\begin{equation}
\label{eq:9}
\begin{aligned}
\rho_{SE}^{(i)}=\sum_{j} a_{ij} \Lambda_{S} (S_{j})=\Lambda_{S}(\rho_{S}^{(i)}).
\end{aligned}
\end{equation}
Now, if $\rho_{SE}^{(i)}\neq \rho_{SE}^{(l)}$, then, at least for one $j$, we have $a_{ij}\neq a_{lj}$. So, from Eq. (\ref{eq:6a}), we conclude that  $\rho_{S}^{(i)}\neq \rho_{S}^{(l)}$. Therefore, there is a one to one  correspondence between the members of  $\mathcal{S}$ and the members of $\mathrm{Tr_{E}}\mathcal{S}$, and so, the $U$-consistency condition is satisfied for the set $\mathcal{S}$, steered from the $\omega_{RSE}$ in Eq. (\ref{eq:2a}).

 In summary, we have proved the following proposition:

\textbf{Proposition 1.} \textit{If the set of possible initial states of the whole system-environment is given by the set of steered states from the  tripartite reference state $\omega_{RSE}$ in Eq. (\ref{eq:2a}), then the reduced dynamics of the system, for arbitrary unitary evolution of the whole system-environment   $U$, is given by a (linear trace-preserving) Hermitian map.}
%$\qquad\qquad\qquad\qquad\qquad\qquad\qquad\qquad\blacksquare$

For the special case that $\Lambda_{S}$ in  Eq. (\ref{eq:2a}) is a CP map, $\omega_{RSE}$ is called a \textit{Markov state} \cite{18}, and the reduced dynamics of the system, for arbitrary  $U$, is, therefore,  CP \cite{15}.
In fact, the reverse is also true. In summary, we have \cite{15}:

\textbf{Theorem 1.}
\textit{ For a set of steered states, from a tripartite reference state $\omega_{RSE}$, as Eq. (\ref{eq:1a}),  the reduced dynamics of the   system, for arbitrary  $U$, is  CP if and only if $\omega_{RSE}$ is a Markov state .}

 \textbf{Remark 1.} \textit{ During the proof of Theorem 1 in Ref.  \cite{15}, it has been assumed that, in general, the dimensions of $\cH_{S}$ and $\cH_{E}$ can vary during the evolution, while the dimension of $\cH_{S}\otimes\cH_{E}$ remains unchanged.}

Interestingly, all the previous results, in this context, are special cases of the above result : All the previously found sets of the system-environment initial states in Refs. \cite{11, 12, 14, 13}, for which the reduced dynamics of the system, for arbitrary  $U$, is  CP, can be written as  steered sets, from  Markov  states $\omega_{RSE}$ \cite{16}.

This fact that  Eq. (\ref{eq:2a}), for the special case of completely positive $\Lambda_{S}$, gives such interesting general results, leads us to this conjecture that  Eq. (\ref{eq:2a}), for the general case of Hermitian $\Lambda_{S}$, can also yield general interesting results. In fact, as we will prove in the   following section, for  arbitrary convex set of  system-environment initial states  $\lbrace\rho_{SE}\rbrace$, which leads to Hermitian reduced dynamics, we can assign a tripartite reference state $\omega_{RSE}$, as Eq. (\ref{eq:2a}).

\section{Reference state for a $U$-consistent subspace} \label{sec:ref state}

Let's denote the convex set of possible initial states of the system-environment as $\mathcal{S}^{\prime}$, and so, the convex set of possible initial states of the system as $\mathcal{S}_{S}^{\prime}=\mathrm{Tr}_{E}\mathcal{S}^{\prime}$.
 Since the Hilbert space of the system $\cH_S$ is finite dimensional, one can find a set $\mathcal{S}_{S}^{\prime\prime}\subset \mathcal{S}^{\prime}_{S}$ including a finite number of $\rho_{S}^{(j)}\in \mathcal{S}^{\prime}_{S}$ which are linearly independent and other states in $\mathcal{S}^{\prime}_{S}$ can be decomposed as linear combinations of them: $\mathcal{S}_{S}^{\prime\prime}=\lbrace\rho_{S}^{(1)},\;\rho_{S}^{(2)},\;\cdots ,\;\rho_{S}^{(m)}\rbrace$, where $m$ is an integer and $m\leq(d_{S})^{2}$  ($d_{S}$ is the dimension of $\cH_S$, so $(d_{S})^{2}$ is the dimension of $\mathcal{L}(\cH_S)$), and, for each $\rho_{S}\in \mathcal{S}^{\prime}_{S}$, we have $\rho_{S}=\sum_{j=1}^{m}b_{j}\,\rho_{S}^{(j)}$ with real $b_{j}$.

Consider the set  $\mathcal{S}^{\prime\prime}=\lbrace\rho_{SE}^{(1)},\;\rho_{SE}^{(2)},\;\cdots ,\;\rho_{SE}^{(m)}\rbrace$, where $\mathrm{Tr}_{E}(\rho_{SE}^{(j)})=\rho_{S}^{(j)}\in \mathcal{S}_{S}^{\prime\prime}$. So, $\rho_{SE}^{(j)}$ are also linearly independent. Now, there is a one to one correspondence between the members of  $\mathcal{S}^{\prime}$ and the members of  $\mathcal{S}_{S}^{\prime}$ if and only if each $\rho_{SE}\in \mathcal{S}^{\prime}$ can be decomposed as a linear combination of $\rho_{SE}^{(j)}\in \mathcal{S}^{\prime\prime}$: $\rho_{SE}=\sum_{j=1}^{m}b_{j}\,\rho_{SE}^{(j)}$. (Note that the coefficients  $b_{j}$ in the decomposition of $\rho_{SE}$ are the same as $b_{j}$ in the decomposition of $\rho_{S}=\mathrm{Tr}_{E}(\rho_{SE})$.)

So, if the set $\mathcal{S}^{\prime\prime}$ constructs a basis for the convex set $\mathcal{S}^{\prime}$, then $\mathcal{S}^{\prime}$ is, in addition, $U$-consistent for arbitrary $U$, and, as we will see in the following, the reduced dynamics of the system is Hermitian.

Now, we can define the linear trace-preserving Hermitian map $\Lambda_{S}$ as $\Lambda_{S}(\rho_{S}^{(j)})=\rho_{SE}^{(j)}$, where $\rho_{S}^{(j)}\in \mathcal{S}_{S}^{\prime\prime}$ and so $\rho_{SE}^{(j)}\in \mathcal{S}^{\prime\prime}$. Therefore, for each $\rho_{S}\in \mathcal{S}_{S}^{\prime}$, we have
\begin{equation}
\label{eq:10}
\begin{aligned}
\Lambda_{S}(\rho_{S})=\sum_{j=1}^{m}b_{j}\, \Lambda_{S}(\rho_{S}^{(j)})=\sum_{j=1}^{m}b_{j}\,\rho_{SE}^{(j)}=\rho_{SE},
\end{aligned}
\end{equation}
where $\rho_{SE}\in \mathcal{S}^{\prime}$ such that $\mathrm{Tr}_{E}(\rho_{SE})=\rho_{S}$. The Hermitian map $\Lambda_{S}$ is called the \textit{assignment map} \cite{3, 21}.
So, from Eq. (\ref{eq:10}), for arbitrary unitary evolution $U$ for the whole system-environment, we have
\begin{equation}
\label{eq:11}
\begin{aligned}
\rho_{S}^{\prime}=\mathrm{Tr_{E}} \circ Ad_{U}(\rho_{SE}) \qquad \qquad\qquad\qquad\quad\\
=\sum_{j=1}^{m}b_{j}\, [\mathrm{Tr_{E}} \circ Ad_{U}\circ \Lambda_{S}](\rho_{S}^{(j)})=\mathcal{E}_{S}(\rho_{S}),
\end{aligned}
\end{equation}
where $\mathcal{E}_{S}=\mathrm{Tr_{E}} \circ Ad_{U}\circ \Lambda_{S}$ is a Hermitian map on $\mathcal{L}(\cH_{S})$, since $\mathrm{Tr_{E}}$ and $Ad_{U}$ are completely positive \cite{1} and $\Lambda_{S}$ is Hermitian.

Now, our  question is as follows: Can we assign to the above convex $U$-consistent $\mathcal{S}^{\prime}$ a tripartite reference state $\omega_{RSE}$, such that $\mathcal{S}^{\prime}$ is the set of steered states from this  $\omega_{RSE}$?

Without loss of generality, as we will show in the following, we consider a restricted set $\mathcal{S}$, instead of $\mathcal{S}^{\prime}$, such that each $\rho_{SE}\in \mathcal{S}$ can be decomposed as $\rho_{SE}=\sum_{l=1}^{m}p_{l}\,\rho_{SE}^{(l)}$, with $\rho_{SE}^{(l)}\in \mathcal{S}^{\prime\prime}$, where $\lbrace p_{l} \rbrace$ is a probability distribution ($p_{l}\geq 0$ and $\sum p_{l}=1$).
As  $\mathcal{S}^{\prime}$, the set $\mathcal{S}$ is  convex (and $U$-consistent) and so the set  $\mathcal{S}_{S}=\mathrm{Tr}_{E}\mathcal{S}$ is also convex.

First, we define the bipartite state
\begin{equation}
\label{eq:12}
\begin{aligned}
\omega_{RS}=\sum_{l=1}^{m} \frac{1}{m} \vert l_{R}\rangle\langle l_{R}\vert\otimes \rho_{S}^{(l)},
\end{aligned}
\end{equation}
where $ \rho_{S}^{(l)}\in \mathcal{S}^{\prime\prime}_{S}$ and $\lbrace \vert l_{R}\rangle\rbrace$ is an orthonormal basis for the reference Hilbert space $\cH_{R}$. So,  using the assignment map $\Lambda_{S}$  in Eq. (\ref{eq:10}), we have
\begin{equation}
\label{eq:13}
\begin{aligned}
\omega_{RSE}\equiv id_{R}\otimes \Lambda_{S} (\omega_{RS}) 
=\sum_{l=1}^{m} \frac{1}{m} \vert l_{R}\rangle\langle l_{R}\vert\otimes \rho_{SE}^{(l)},
\end{aligned}
\end{equation}
where $ \rho_{SE}^{(l)}\in \mathcal{S}^{\prime\prime}$ such that $ \mathrm{Tr}_{E}( \rho_{SE}^{(l)})=\rho_{S}^{(l)}$. Therefore, using Eq. (\ref{eq:1a}), we can write the set $\mathcal{S}$ as the steered set from the tripartite reference state $\omega_{RSE}$, given in Eq.  (\ref{eq:13}). It can be done, e.g.,  by  considering the positive operators $P_{R}$ in Eq. (\ref{eq:1a}) as $P_{R}=\sum_{l=1}^{m} mp_{l}\vert l_{R}\rangle\langle l_{R}\vert$.
 Note that $\omega_{RSE}$, in Eq. (\ref{eq:13}), is in the form of Eq. (\ref{eq:2a}), with the Hermitian assignment map $\Lambda_{S}$. 

In summary, we have proved the following theorem:

\textbf{Theorem 2.} \textit{Consider a set of linearly independent states $\mathcal{S}_{S}^{\prime\prime}=\lbrace\rho_{S}^{(1)},\;\rho_{S}^{(2)},\;\cdots ,\;\rho_{S}^{(m)}\rbrace$. So, the set $\mathcal{S}^{\prime\prime}=\lbrace\rho_{SE}^{(1)},\;\rho_{SE}^{(2)},\;\cdots ,\;\rho_{SE}^{(m)}\rbrace$,  such that $\mathrm{Tr}_{E}(\rho_{SE}^{(l)})=\rho_{S}^{(l)}$  (for each $1\leq l\leq m$), is also linearly independent. The set $\mathcal{S}$ of the convex combinations of  $\rho_{SE}^{(l)}\in \mathcal{S}^{\prime\prime}$ is convex and $U$-consistent, for arbitrary unitary evolution of the whole system-environment. Therefore, if the set of possible initial states of the system-environment is given by $\mathcal{S}$, then the reduced dynamics of the system is given by a Hermitian map $\mathcal{E}_{S}$. In addition, $\mathcal{S}$ can be written as the steered set from a tripartite reference state $\omega_{RSE}$, given in Eq.  (\ref{eq:13}), which is in the form of Eq. (\ref{eq:2a}), with the Hermitian assignment map $\Lambda_{S}$.}
%$\qquad\qquad\qquad\qquad\qquad\qquad\blacksquare$

Next,  let's define $\mathcal{V}\subseteq \mathcal{L}(\cH_{S}\otimes\cH_{E})$ as the subspace spanned by the states $\rho_{SE}^{(l)}\in \mathcal{S}^{\prime\prime}$; i.e., for each $X\in \mathcal{V}$, we have $X=\sum_{l}c_{l}\,\rho_{SE}^{(l)}$ with unique complex coefficients $c_{l}$. Obviously $\mathcal{S}\subseteq  \mathcal{S}^{\prime} \subset\mathcal{V}$.

So, the subspace $\mathcal{V}_{S}=\mathrm{Tr}_{E}\mathcal{V}\subseteq \mathcal{L}(\cH_{S})$ is spanned by the states $\rho_{S}^{(l)}\in \mathcal{S}_{S}^{\prime\prime}$: For each $X\in \mathcal{V}$, we have $x=\mathrm{Tr}_{E}(X)=\sum_{l}c_{l}\,\rho_{S}^{(l)}$ with the same coefficients $c_{l}$ as in the decomposition of $X$.

Note that, since there is a one to one correspondence between the $x\in \mathcal{V}_{S}$ and the $X\in \mathcal{V}$, the whole subspace  $\mathcal{V}$ is $U$-consistent, for arbitrary $U$. 

In addition, we can write the subspace  $\mathcal{V}$  as
\begin{equation}
\label{eq:14}
\begin{aligned}
\mathcal{V}=\left\lbrace \mathrm{Tr_{R}}[(A_{R}\otimes I_{SE})\omega_{RSE}] ,  A_{R}\in \mathcal{L}(\cH_{R}) \right\rbrace ,  
\end{aligned}
\end{equation}
where $ A_{R}$ is arbitrary linear operator in $\mathcal{L}(\cH_{R})$ and $\omega_{RSE}$ is the reference state, given in Eq.  (\ref{eq:13}). We will call the above set the \textit{generalized steered} set from the reference state $\omega_{RSE}$. Using this fact that if the subspace  $\mathcal{V}$ is $U$-consistent, for arbitrary $U$, then there is a one to one correspondence between the $x\in \mathcal{V}_{S}$ and the $X\in \mathcal{V}$ \cite{3}, we can write the above result in the following form:

\textbf{Corollary 1.} \textit{Consider the subspace $\mathcal{V}\subseteq \mathcal{L}(\cH_{S}\otimes\cH_{E})$, which is spanned by states. If $\mathcal{V}$ is $U$-consistent, for arbitrary $U$, then it can be written as the generalized steered set from the reference state $\omega_{RSE}$, as  Eq.  (\ref{eq:14}).}
%$\quad \blacksquare$

Note that, since $\mathcal{S}^{\prime} \subset\mathcal{V}$, $\mathcal{S}^{\prime}$ can be written as (a subset of) Eq.  (\ref{eq:14}). In our discussion, leading to the reference state $\omega_{RSE}$ in Eq.  (\ref{eq:13}), we have restricted ourselves to the set $\mathcal{S}$, instead of  $\mathcal{S}^{\prime}$. Now, 
 as stated before, we see that this restriction does not lose the generality of our discussion.

The next observation is that the evolution of the system subspace $\mathcal{V}_{S}$ and the whole system-environment subspace  $\mathcal{V}$ can be given from the evolution of $\omega_{RS}$ in Eq.  (\ref{eq:12}), and $\omega_{RSE}$ in Eq.  (\ref{eq:13}), respectively. So, we can call $\omega_{RS}$ as the reference state of the system and $\omega_{RSE}$ as the reference state of the whole system-environment. Note that these two reference states are related to each other as Eq. (\ref{eq:2a}).

Assume that the unitary time evolution of the whole system-environment, from the initial instant to the time $t$, is given by $U(t)$. So, $\omega_{RSE}$  evolves as
\begin{equation}
\label{eq:15}
\begin{aligned}
\omega_{RSE}(t)= id_{R}\otimes Ad_{U(t)} (\omega_{RSE}(0)),
\end{aligned}
\end{equation}
where $\omega_{RSE}(0)$ is given in Eq.  (\ref{eq:13}). As stated before, each $X\in \mathcal{V}=\mathcal{V}(0)$ can be written as $X=X(0)=\sum_{l}c_{l}\,\rho_{SE}^{(l)}=\mathrm{Tr_{R}}[(A_{R}\otimes I_{SE})\omega_{RSE}(0)]$. So, $X(t)=\sum_{l}c_{l}\,Ad_{U(t)}(\rho_{SE}^{(l)})$ and therefore
\begin{equation}
\label{eq:16}
\begin{aligned}
\mathcal{V}(t)=\lbrace X(t)\rbrace=\left\lbrace \mathrm{Tr_{R}}[(A_{R}\otimes I_{SE})\omega_{RSE}(t)] \right\rbrace ,  
\end{aligned}
\end{equation}
where $ A_{R}$ is arbitrary linear operator in $\mathcal{L}(\cH_{R})$ and $\omega_{RSE}(t)$ is the reference state of the system-environment, given in Eq.  (\ref{eq:15}).

Similarly, $\omega_{RS}$  evolves as
\begin{equation}
\label{eq:17}
\begin{aligned}
\omega_{RS}(t)= id_{R}\otimes \mathcal{E}_{S}(t) (\omega_{RS}(0)),
\end{aligned}
\end{equation}
where $\omega_{RS}(0)$ is given in Eq.  (\ref{eq:12}) and $\mathcal{E}_{S}(t)=\mathrm{Tr}_{E}\circ Ad_{U(t)}\circ \Lambda_{S}$ is a Hermitian map on $\mathcal{L}(\cH_{S})$. Each $x=x(0)=\mathrm{Tr}_{E}(X(0))$ can be decomposed as $x(0)=\sum_{l}c_{l}\,\rho_{S}^{(l)}$. So, $x(t)=\sum_{l}c_{l}\,\mathcal{E}_{S}(t)(\rho_{S}^{(l)})=\mathcal{E}_{S}(t)(x(0))$ and therefore
\begin{equation}
\label{eq:18}
\begin{aligned}
\mathcal{V}_{S}(t)=\lbrace x(t)\rbrace=\left\lbrace \mathrm{Tr_{R}}[(A_{R}\otimes I_{S})\omega_{RS}(t)] \right\rbrace ,  
\end{aligned}
\end{equation}
where $ A_{R}$ is arbitrary linear operator in $\mathcal{L}(\cH_{R})$ and $\omega_{RS}(t)$ is the reference state of the system, given in Eq.  (\ref{eq:17}).

In summary,

\textbf{Corollary 2.} \textit{Consider the subspace $\mathcal{V}(0)\subseteq \mathcal{L}(\cH_{S}\otimes\cH_{E})$, which is spanned by states. If $\mathcal{V}(0)$ is $U$-consistent, for arbitrary $U$, then $\mathcal{V}(t)$ and $\mathcal{V}_{S}(t)$   can be written as the generalized steered sets, from the reference states $\omega_{RSE}(t)$, in  Eq.  (\ref{eq:15}), and $\omega_{RS}(t)$, in  Eq.  (\ref{eq:17}), respectively.}
%$\qquad\qquad\qquad\qquad\qquad\qquad\qquad\qquad\qquad\qquad\blacksquare$

In general, there are more than one possible assignment maps $\Lambda_{S}$. So, it may be possible, by choosing an appropriate $\Lambda_{S}$, to write the reduced dynamics of the system $S$ as a CP map. Note that, from Eqs. 
(\ref{eq:12}) and (\ref{eq:17}), we have 
\begin{equation}
\label{eq:12a}
\begin{aligned}
\omega_{RS}(t)=\sum_{l=1}^{m} \frac{1}{m} \vert l_{R}\rangle\langle l_{R}\vert\otimes \rho_{S}^{(l)}(t).
\end{aligned}
\end{equation}
Now, if the time evolution of the system can be written as a CP map, then, since $\rho_{S}^{(l)}(t)= \mathcal{E}_{S}^{(CP)}(t)(\rho_{S}^{(l)})$, where $\mathcal{E}_{S}^{(CP)}(t)$ is a CP map on $\mathcal{L}(\cH_{S})$, we have
\begin{equation}
\label{eq:12b}
 \omega_{RS}(t)= id_{R}\otimes \mathcal{E}_{S}^{(CP)}(t) (\omega_{RS}(0)),
\end{equation}
i.e., even if we have used a  $\Lambda_{S}$ which leads to Eq. (\ref{eq:17}), with a non-CP map $\mathcal{E}_{S}(t)$, $\omega_{RS}(t)$ can be written as Eq. (\ref{eq:12b}), too.

 Reversely,
if $\omega_{RS}(t)$, in Eq. (\ref{eq:12a}), can be written as  Eq. (\ref{eq:12b}), then, using this fact that  $m\langle l_{R} \vert \omega_{RS}(t) \vert l_{R} \rangle=\rho_{S}^{(l)}(t)$, we, simply, conclude that  $\rho_{S}^{(l)}(t)= \mathcal{E}_{S}^{(CP)}(t)(\rho_{S}^{(l)})$, and so, the reduced dynamics of the system is given by the CP map $\mathcal{E}_{S}^{(CP)}(t)$. In summary,

\textbf{Theorem 3.} \textit{The reduced dynamics of the system can be written as a CP map if and only if the reference state  $\omega_{RS}(t)$, in Eq. (\ref{eq:12a}), evolves as Eq. (\ref{eq:12b}), with a CP map $\mathcal{E}_{S}^{(CP)}(t)$.}

\section{Markovianity of the reference state and the complete positivity of the reduced dynamics} \label{sec:CP}

Theorem 1 states the relation between the Markovianity of the reference state $\omega_{RSE}$ and the CP-ness of the reduced dynamics, 
 for a steered set as Eq. (\ref{eq:1a}).
 %, the reduced dynamics of the system $S$ is CP, for arbitrary $U$, if and only if, the reference state $\omega_{RSE}$ is a Markov state. 
In the previous section, we have seen that, for an arbitrary $U$-consistent subspace  $\mathcal{V}$, we can also introduce a reference state as Eq. (\ref{eq:13}), such that $\mathcal{V}$ can be written as the generalized steered set from it.
 Therefore, we conjecture that Theorem 1 can be generalized to arbitrary $U$-consistent subspace  $\mathcal{V}$. Fortunately, this is the case, as we will show in this section.

%As stated before in Sec. ~\ref{sec:steered set}, 
A tripartite state $\rho_{RSE}$ is called a Markov state if it can be written as $\rho_{RSE}=id_{R}\otimes \bar{\Lambda}_{S} (\rho_{RS})$, where $\rho_{RS}= \mathrm{Tr_{E}}(\rho_{RSE})$, and  $\bar{\Lambda}_{S}\,:\mathcal{L}(\cH_{S})\rightarrow\mathcal{L}(\cH_{S}\otimes\cH_{E})$ is a CP assignment map  \cite{18}.
Now, it has been shown
in Ref. \cite{18}  that if $\rho_{RSE}$  is  a Markov state, then  there exists a decomposition of the Hilbert space of the system $S$ as $\cH_{S}=\bigoplus_{k}\cH_{s_{k}}=\bigoplus_{k}\cH_{s^{l}_{k}}\otimes\cH_{s^{r}_{k}}$  such that
\begin{equation}
\label{eq:33}
\rho_{RSE}=\bigoplus_{k}\lambda_{k}\:\rho_{Rs^{l}_{k}}\otimes\rho_{s^{r}_{k}E},
\end{equation}
where $\lbrace \lambda_{k}\rbrace$ is a probability distribution, $\rho_{Rs^{l}_{k}}$ is a state on $\cH_{R}\otimes\cH_{s^{l}_{k}}$, and $\rho_{s^{r}_{k}E}$ is a state on $\cH_{s^{r}_{k}}\otimes\cH_{E}$.

Consider a set  $\mathcal{S}^{\prime\prime}$ which spans the subspace  $\mathcal{V}$.
% (and so  fixed $\mathcal{S}_{S}^{\prime\prime}$, $\omega_{RSE}$ and $\omega_{RS}$),
 In general, we can consider different assignment maps $\Lambda_{S}$ such that, for all of them, $\Lambda_{S}(\rho_{S}^{(j)})=\rho_{SE}^{(j)}$, where $\rho_{S}^{(j)}\in \mathcal{S}_{S}^{\prime\prime}$ and  $\rho_{SE}^{(j)}\in \mathcal{S}^{\prime\prime}$, and so, we can write Eq. (\ref{eq:10}), for all of them.
Different assignment maps $\Lambda_{S}$ can lead to different reduced dynamics $\mathcal{E}_{S}$, in Eq. 
(\ref{eq:11}).

Therefore, it is possible that we choose a Hermitian (non-CP) assignment map $\Lambda_{S}$ to construct the 
 reference state  $\omega_{RSE}$ in Eq. (\ref{eq:13}), while there is another CP assignment map  which could be used instead. So, the reduced dynamics could be written as a CP map, while we write it as a non-CP map. How can we avoid such inappropriate choosing?
 
Note that if there is a CP assignment map $\bar{\Lambda}_{S}$, such that, for all $\rho_{S}^{(j)}\in \mathcal{S}_{S}^{\prime\prime}$, we have $\bar{\Lambda}_{S}(\rho_{S}^{(j)})=\rho_{SE}^{(j)}\in \mathcal{S}^{\prime\prime}$, then  $\omega_{RSE}$ in Eq. (\ref{eq:13}) is a Markov state, even if we have used a 
non-CP assignment map $\Lambda_{S}$ to construct it.
So, we can check whether  $\omega_{RSE}$ can be written as Eq.  (\ref{eq:33}), or not. If it can be written so, then the   reference state  $\omega_{RSE}$ is a Markov state, and the reduced dynamics is CP, for arbitrary $U$.

 But if  $\omega_{RSE}$ cannot be written as Eq.  (\ref{eq:33}), then we conclude that there is no CP assignment map which can map all $\rho_{S}^{(j)}\in \mathcal{S}_{S}^{\prime\prime}$ to $\rho_{SE}^{(j)}\in \mathcal{S}^{\prime\prime}$. In other words, though there may be more than one possible assignment maps $\Lambda_{S}$, but none of them is CP.
 
Also note that,   for the subspace  $\mathcal{V}$, we can construct different reference states  $\omega_{RSE}$ as Eq. (\ref{eq:13}), in general: By choosing a different set  $\mathcal{S}^{\prime\prime}$, which also spans  $\mathcal{V}$, we can construct a different $\omega_{RSE}$.
Interestingly, if the previously constructed reference state is non-Markovian,  this new reference state is not a Markov state, too; otherwise, there is a CP assignment map which maps all $\rho_{S}\in \mathcal{V}_{S}$ to $\rho_{SE}\in \mathcal{V}$.

In summary, we have proved the following theorem:

\textbf{Theorem 4.} \textit{Consider the  subspace $\mathcal{V}\subseteq \mathcal{L}(\cH_{S}\otimes\cH_{E})$, which is spanned by states and is $U$-consistent, for arbitrary $U$. One can find, at least, one CP assignment map $\bar{\Lambda}_{S}\,:\mathcal{L}(\cH_{S})\rightarrow\mathcal{L}(\cH_{S}\otimes\cH_{E})$, which maps $\mathcal{V}_{S}=\mathrm{Tr_{E}}\mathcal{V}$ to $\mathcal{V}$, if and only if any reference state  
 $\omega_{RSE}$, which is constructed as Eq. (\ref{eq:13}), is a Markov state, as Eq. (\ref{eq:33}).}

 Next, consider the case that the reference state $\omega_{RSE}$, in Eq. (\ref{eq:13}), is not a Markov state.  
 Construct   the set of 
  steered states from  $\omega_{RSE}$, i.e., the set $\mathcal{S}$ in Theorem 2.
  % In Ref. \cite{15}, it has been shown that, for a steered set of initial states of the system-environment, the reference state $\omega_{RSE}$ is a Markov state if and only if the reduced dynamics of the system $S$ is CP, for arbitrary $U$. Therefore, from the non-Markovianity of $\omega_{RSE}$,
 Using Theorem 1, we conclude that the reduced dynamics of the system is non-CP, for at least one $U$. Since $\mathcal{S}\subset\mathcal{V}$, the non-CP-ness of the reduced dynamics for $\mathcal{S}$ results in the non-CP-ness of the reduced dynamics for $\mathcal{V}$. In other words,
 
\textbf{ Theorem 5.} \textit{Consider the  subspace $\mathcal{V}\subseteq \mathcal{L}(\cH_{S}\otimes\cH_{E})$, which is $U$-consistent, for arbitrary $U$, and can be written as the generalized steered set, from 
 the reference state $\omega_{RSE}$, in Eq. (\ref{eq:13}). When $\omega_{RSE}$ is not a Markov state as Eq. (\ref{eq:33}), then the reduced dynamics of the system, for at least one $U$, is non-CP.}

 The above theorem, states that, not only for a steered set of initial states of the system-environment as Eq. (\ref{eq:1a}), but also, for any arbitrary  subspace $\mathcal{V}$, which is $U$-consistent for all $U$, the reduced dynamics of the system, for arbitrary $U$, is CP if and only if the reference state $\omega_{RSE}$ is a Markov state. This is the generalization of  Theorem 1, to arbitrary $U$-consistent subspace $\mathcal{V}$.
 
The following point is also worth noting:

\textbf{Corollary 3.} \textit{ Theorems 4 and 5 state that the impossibility of a CP assignment map is equivalent to non-CP-ness of the reduced dynamics, for at least one $U$.}

\section{Generalization to arbitrary $\mathcal{G}$-consistent subspace} \label{sec:generalization}
%Generalization to arbitrary $\mathcal{G}$-consistent subspace

Till now, our discussion was restricted to the case that there is a one to one correspondence between the members of $\mathcal{V}$ and   $\mathcal{V}_{S}$.
We can generalize our discussion to include the general case (with no such correspondence), straightforwardly.

Consider the subspace $\mathcal{V}\subseteq \mathcal{L}(\cH_{S}\otimes\cH_{E})$, which is spanned by states. If there is not a one to one correspondence between the members of $\mathcal{V}$ and the members of  $\mathcal{V}_{S}=\mathrm{Tr}_{E}\mathcal{V}$, then $\mathcal{V}$ is $U$-consistent only for a restricted set  $\mathcal{G}\subset\mathcal{U}(\cH_{S}\otimes\cH_{E})$ ($\mathcal{U}(\cH_{S}\otimes\cH_{E})$, is the set of all unitary $U\in\mathcal{L}(\cH_{S}\otimes\cH_{E})$) \cite{3}.
In such case, the subspace 
 $\mathcal{V}$ is called a $\mathcal{G}$-consistent subspace.

Assume that the set of linearly independent states $\mathcal{S}^{\prime\prime}=\lbrace\rho_{SE}^{(1)},\;\rho_{SE}^{(2)},\;\cdots ,\;\rho_{SE}^{(M)}\rbrace$, where $M$ is an integer  such that $ M\leq (d_{S}d_{E})^2$   ($d_{S}$  and $d_{E}$ are the dimensions of $\cH_S$ and  $\cH_E$, respectively), spans the subspace 
 $\mathcal{V}$.
Without loss of generality, we can assume that only $\rho_{S}^{(l)}=\mathrm{Tr}_{E}(\rho_{SE}^{(l)})$, for $1\leq l\leq m$  (where the integer $m\leq (d_{S})^{2}$ is, in addition, less than $M$), are linearly independent. So, the subspace  $\mathcal{V}_{S}$ is spanned by the set of states $\mathcal{S}_{S}^{\prime\prime}=\lbrace\rho_{S}^{(1)},\;\rho_{S}^{(2)},\;\cdots ,\;\rho_{S}^{(m)}\rbrace$.

As before, we can define the (linear trace-preserving) Hermitian assignment map $\Lambda_{S}$ as $\Lambda_{S}(\rho_{S}^{(j)})=\rho_{SE}^{(j)}$, where $\rho_{S}^{(j)}\in \mathcal{S}_{S}^{\prime\prime}$,   $\rho_{SE}^{(j)}\in \mathcal{S}^{\prime\prime}$ and $1\leq j\leq m$, and so, we can write a similar relation as Eq.  (\ref{eq:10}), for each $x\in \mathcal{V}_{S}$. Therefore, the assignment map $\Lambda_{S}$ maps $\mathcal{V}_{S}$ to a subspace   $\mathcal{V}^{\prime}\subset\mathcal{V}$, which is spanned by $\lbrace\rho_{SE}^{(1)},\;\rho_{SE}^{(2)},\;\cdots ,\;\rho_{SE}^{(m)}\rbrace$.

Note that 
\begin{equation}
\label{eq:19}
\begin{aligned}
\mathcal{V}=\mathcal{V}^{\prime}\oplus\mathcal{V}_{0},
\end{aligned}
\end{equation}
where, for each $Y\in \mathcal{V}_{0}$, we have $\mathrm{Tr}_{E}(Y)=0$. So, the most general possible assignment map is as
\begin{equation}
\label{eq:20}
\begin{aligned}
\tilde{\Lambda}_{S}=\Lambda_{S}+\mathcal{V}_{0},
\end{aligned}
\end{equation}
where  $\mathcal{V}_{0}$ denotes   arbitrary  $Y\in \mathcal{V}_{0}$. 

Each $U\in \mathcal{G}$ maps $\mathcal{V}_{0}$ to $\mathrm{kerTr}_{E}$, the set of all $Z\in \mathcal{L}(\cH_{S}\otimes\cH_{E})$ for which we have $\mathrm{Tr}_{E}(Z)=0$, and so, $\mathcal{V}$ is $U$-consistent under all $U\in \mathcal{G}$ \cite{3}. Therefore, for a unitary time evolution $U(t)\in \mathcal{G}$, from Eq. (\ref{eq:19}), we have
\begin{equation}
\label{eq:21}
\begin{aligned}
\mathcal{V}(t)=\mathcal{V}^{\prime}(t)\oplus\mathcal{V}_{0}(t),
\end{aligned}
\end{equation}
where $\mathcal{V}_{0}(t)\subseteq \mathrm{kerTr}_{E}$ and $\mathcal{V}^{\prime}(t)$ is given as Eq. (\ref{eq:16}), i.e. as the generalized steered set from the reference state $\omega_{RSE}(t)$ in Eq.  (\ref{eq:15}).

In addition, for each $x=x(0)\in \mathcal{V}_{S}=\mathcal{V}_{S}(0)$ 
and each $U(t)\in \mathcal{G}$, 
we have $x(t)=[\mathrm{Tr}_{E}\circ Ad_{U(t)}\circ \tilde{\Lambda}_{S}](x)=[\mathrm{Tr}_{E}\circ Ad_{U(t)}\circ \Lambda_{S}](x)=\mathcal{E}_{S}(t)(x)$. Therefore, as before, $\mathcal{V}_{S}(t)$ can be written as Eq. (\ref{eq:18}), i.e. as the generalized steered set from the reference state $\omega_{RS}(t)$ in Eq.  (\ref{eq:17}).

So, we have proved the following proposition:

\textbf{Proposition 2.}
\textit{Consider the $\mathcal{G}$-consistent  subspace $\mathcal{V}(0)\subseteq \mathcal{L}(\cH_{S}\otimes\cH_{E})$, which is spanned by states. For each $U(t)\in \mathcal{G}$,   $\mathcal{V}_{S}(t)$, even simply,  can be written as the generalized steered set,  from the reference state $\omega_{RS}(t)$ in  Eq.  (\ref{eq:17}). In addition, $\mathcal{V}(t)=\mathcal{V}^{\prime}(t)\oplus\mathcal{V}_{0}(t)$, where $\mathcal{V}_{0}(t)$ is a subset of $\mathrm{kerTr}_{E}$ and $\mathcal{V}^{\prime}(t)$ can be written as the generalized steered set,  from the reference state $\omega_{RSE}(t)$ in  Eq.  (\ref{eq:15}).}
%$\qquad\qquad\qquad\qquad\qquad\qquad\qquad\qquad\qquad\qquad\quad\blacksquare$

Note that Theorem 3 is  valid for a $\mathcal{G}$-consistent  subspace, too, since, even simply, the reduced dynamics of the system is determined by the evolution of the reference state  $\omega_{RS}(t)$.

In the following, we discuss about the generalization of the results given in the previous section, to a 
 $\mathcal{G}$-consistent  subspace $\mathcal{V}$. First, Theorem 4 is changed as below:

\textbf{Theorem $4^{\prime}$.} \textit{Consider a  $\mathcal{G}$-consistent  subspace $\mathcal{V}$, which is spanned by states. There exists, at least, one CP assignment map $\bar{\Lambda}_{S}$ if and only if,  at least, one reference state $\omega_{RSE}$, as  Eq.  (\ref{eq:13}), is a Markov state, as Eq. (\ref{eq:33}).}
% such that $\mathcal{V}^{\prime}$ can be written as the generalized steered set from it.}

Note that when there exists a  CP assignment map $\bar{\Lambda}_{S}$, then using this $\bar{\Lambda}_{S}$ in Eq. (\ref{eq:13}), we can construct a Markov reference state  $\omega_{RSE}$.
But, from the CP-ness of $\bar{\Lambda}_{S}$, we cannot, in general, conclude that  $\tilde{\Lambda}_{S}=\bar{\Lambda}_{S}+\mathcal{V}_{0}$ is also CP. So, in general, one can construct other reference states which are not Markov states.
However, if, for our $\mathcal{G}$-consistent  subspace $\mathcal{V}$, we can find a reference state 
 $\omega_{RSE}$, as Eq. (\ref{eq:13}), which is a Markov state, as Eq. (\ref{eq:33}), then the reduced dynamics of the system is CP, for any arbitrary $U\in \mathcal{G}$.
 %This case, for which there exists a CP assignment map, has been studied in Ref. \cite{25}.

Unfortunately,
 Theorem 5 cannot be generalized to a  $\mathcal{G}$-consistent  subspace $\mathcal{V}$, in general. Assume that the reduced dynamics of the system $\mathcal{E}_{S}$ is CP, for any arbitrary $U\in \mathcal{G}$. The CP-ness of the reduced dynamics, for any $\rho_{SE}\in \mathcal{V}$, results in  the CP-ness of the reduced dynamics, for any convex set of initial states $\mathcal{S}=\lbrace\rho_{SE}\rbrace\subset\mathcal{V}$. Therefore, for the steered set $\mathcal{S}$, from any reference state $\omega_{RSE}$, constructed as Eq. (\ref{eq:13}), the reduced dynamics is CP, for any  arbitrary $U\in \mathcal{G}$. But, from this result,  we cannot (in general) conclude that $\omega_{RSE}$ is a Markov state, unless  $\mathcal{G}=\mathcal{U}(\cH_{S}\otimes\cH_{E})$, which is the case considered in the previous section. In fact, as we will see in the next section, the reduced dynamics can be CP, for some (but not all) $U$, even though $\omega_{RSE}$ is not a Markov state.

%  So, any other  reference state 
% $\omega_{RSE}$ of $\mathcal{V}$, which is constructed as (\ref{eq:13}), is also a Markov state. In Summary:
 
%\textbf{ Theorem 4.} Consider the $\mathcal{G}$-consistent subspace $\mathcal{V}\subseteq \mathcal{L}(\cH_{S}\otimes\cH_{E})$, which is spanned by states. The reduced dynamics of the sy

\section{Example}\label{sec:example}

In Ref. \cite{5}, a two-qubit case, one as the system $S$ and the other as the environment $E$, has been  considered.
First, note that an arbitrary  state of the system can be written as
\begin{equation}
\label{eq:22}
\begin{aligned}
\rho_{S}=\frac{1}{2}(I_{S}+ \vec{\alpha}.\vec{\sigma}_{S}),
\end{aligned}
\end{equation}
where $\vec{\sigma}_{S}=(\sigma^{(1)}_{S}, \sigma^{(2)}_{S}, \sigma^{(3)}_{S})$, $\sigma^{(i)}_{S}$ are the Pauli operators, and the \textit{Bloch vector} $\vec{\alpha}=(\alpha^{(1)}, \alpha^{(2)}, \alpha^{(3)})$ is a real
 three dimensional
 vector such that $\vert\vec{\alpha}\vert\leq 1$ \cite{1}.

Consider the following (linear trace-preserving) Hermitian assignment map $\Lambda_{S}$:
\begin{equation}
\label{eq:23}
\begin{aligned}
\Lambda_{S}(\sigma_{S}^{(i)})=\frac{1}{2}\sigma_{S}^{(i)}\otimes I_{E}\equiv X^{(i)} \qquad  (i=1,\, 2,\,  3),\\
\Lambda_{S}(I_{S})=\frac{1}{2}\left( I_{SE}+a\sum_{i=1}^{3} \sigma^{(i)}_{S}\otimes \sigma^{(i)}_{E}\right)\equiv X^{(4)},
\end{aligned}
\end{equation}
where $a$ is a fixed real constant. For the special case that $a=0$, we have  $\Lambda_{S}(x)=x\otimes (\frac{1}{2}I_{E})$, for each $x\in \mathcal{L}(\cH_{S})$, i.e., $\Lambda_{S}$ is a CP map, in the form first introduced by Pechukas \cite{21, 8}. We denote this special case of  $\Lambda_{S}$ as  $\Lambda^{(CP)}_{S}$. But, for $a\neq 0$,  $\Lambda_{S}$ is not CP. We have
\begin{equation}
\label{eq:24}
\begin{aligned}
\tau_{SE}\equiv\Lambda_{S}(\rho_{S})\qquad\qquad\qquad\qquad\qquad\qquad\qquad\qquad\qquad\\
=\frac{1}{4}\left(I_{SE}+\sum_{i=1}^{3}\alpha^{(i)}\sigma_{S}^{(i)}\otimes I_{E}
 +a\sum_{i=1}^{3} \sigma^{(i)}_{S}\otimes \sigma^{(i)}_{E}\right).
\end{aligned}
\end{equation}
When $a\geq 0$, $\tau_{SE}$ is positive for $\vert\vec{\alpha}\vert\leq \sqrt{(1+a)(1-3a)}$, and when $a\leq 0$,  $\tau_{SE}$ is positive for $\vert\vec{\alpha}\vert\leq (1+a)$ \cite{3, 5}. Therefore, for $a\neq 0$,  $\Lambda_{S}$ is not even a positive map and, consequently, it is not a CP map.

Within the positivity domain of $\tau_{SE}$, i.e., $-1<a< \frac{1}{3}$, we can
apply the framework of Ref. \cite{3}. we can construct $\mathcal{V}$ as \cite{3}
\begin{equation}
\label{eq:25}
\begin{aligned}
\mathcal{V}=\mathrm{Span}_{\mathbb{C}}\lbrace X^{(i)}\rbrace,
\end{aligned}
\end{equation}
i.e., each $X \in \mathcal{V}$ can be decomposed as $X=\sum_{i=1}^{4}c_{i}X^{(i)}$, with complex coefficients $c_{i}$.
(Out of the positivity domain, $\tau_{SE}$, in Eq. (\ref{eq:24}), is not a state. In other words, $\mathcal{V}$ does not contain any state, and so, is not spanned by states.)
 Therefore,
\begin{equation}
\label{eq:26}
\begin{aligned}
\mathcal{V}_{S}=\mathrm{Span}_{\mathbb{C}}\lbrace \sigma^{(1)}_{S}, \sigma^{(2)}_{S}, \sigma^{(3)}_{S}, I_{S} \rbrace =\mathcal{L}(\cH_{S}).
\end{aligned}
\end{equation}
From Eqs. (\ref{eq:25}) and (\ref{eq:26}), we see that there is a one to one correspondence  between the members of $\mathcal{V}$ and the members of  $\mathcal{V}_{S}$. Therefore,  $\mathcal{V}$ is a $U$-consistent subspace, for arbitrary unitary evolution  of the whole system-environment $U$,
and so, the reduced dynamics of the system, from Eq. (\ref{eq:two}) (when $\tau_{SE}$, in Eq. (\ref{eq:24}),  is positive), is given by
\begin{equation}
\label{eq:27}
\begin{aligned}
\rho_{S}^{\prime}=\mathrm{Tr_{E}} \circ Ad_{U}(\tau_{SE})=\mathrm{Tr_{E}}\circ Ad_{U} \circ\Lambda_{S}(\rho_{S})=\mathcal{E}_{S}(\rho_{S}).
\end{aligned}
\end{equation} 

Since $\Lambda_{S}$ is Hermitian (and not CP), we expect that the reduced dynamics $\mathcal{E}_{S}$ be so, in general. But, interestingly, when $U$ commutes with $\sum \sigma^{(i)}_{S}\otimes \sigma^{(i)}_{E}$, $\mathcal{E}_{S}$ is CP  \cite{5}. For  such $U$, we have
\begin{equation}
\label{eq:28}
\begin{aligned}
\mathcal{E}_{S}=\mathrm{Tr_{E}} \circ Ad_{U} \circ\Lambda_{S}=\mathrm{Tr_{E}} \circ Ad_{U} \circ\Lambda_{S}^{(CP)},
\end{aligned}
\end{equation} 
which is a CP map. An interesting question is  whether this result can be generalized to other $U$ or
%, for the assignment map $\Lambda_{S}$, in Eq. (\ref{eq:23}),
 we can  find, at least, one $U$, for which the reduced dynamics $\mathcal{E}_{S}$ is not CP.

This question can be answered simply, using Theorem 5.
% the reference state $\omega_{RSE}$, introduced in Sec. ~\ref{sec:ref state}. 
For an $a$ within the positivity domain $-1<a< \frac{1}{3}$, we, first, choose four states $\rho_{S}^{(l)}$, which  can span $\mathcal{V}_{S}$:
\begin{equation}
\label{eq:29}
\begin{aligned}
\rho_{S}^{(l)}=\frac{1}{2}(I_{S}+ \alpha^{(l)}\sigma^{(l)}_{S})  \qquad (l=1,\, 2,\, 3), \\
\rho_{S}^{(4)}=\frac{1}{2}I_{S}, \qquad\qquad\qquad\qquad\qquad\qquad\,
\end{aligned}
\end{equation}
where $\alpha^{(l)}$ is an arbitrary real constant such that, for $a\geq 0$,  $0< \vert \alpha^{(l)}\vert\leq \sqrt{(1+a)(1-3a)}$, and for $a\leq 0$,  $0< \vert\alpha^{(l)}\vert\leq (1+a)$.
Therefore, from Eq.  (\ref{eq:12}), we can construct the reference state $\omega_{RS}$ as
\begin{equation}
\label{eq:30}
\begin{aligned}
\omega_{RS}=\sum_{l=1}^{4} \frac{1}{4} \vert l_{R}\rangle\langle l_{R}\vert\otimes \rho_{S}^{(l)} \qquad\qquad\qquad\qquad\qquad\qquad\quad\\
=\sum_{l=1}^{3} \frac{1}{8} \vert l_{R}\rangle\langle l_{R}\vert\otimes (I_{S}+ \alpha^{(l)}\sigma^{(l)}_{S})
+ \frac{1}{8} \vert 4_{R}\rangle\langle 4_{R}\vert\otimes I_{S}. 
\end{aligned}
\end{equation}

Next, using Eqs.  (\ref{eq:24}) and  (\ref{eq:29}), we can construct four states $\rho_{SE}^{(l)}=\Lambda_{S}(\rho_{S}^{(l)})$, which  span $\mathcal{V}$:
\begin{equation}
\label{eq:31}
\begin{aligned}
\rho_{SE}^{(l)}
=\frac{1}{4}(I_{SE}+ \alpha^{(l)}\sigma_{S}^{(l)}\otimes I_{E} \qquad\qquad\qquad\qquad \\
 +a\sum_{i=1}^{3} \sigma^{(i)}_{S}\otimes \sigma^{(i)}_{E}),\qquad (l=1,2,3), \\
 \rho_{SE}^{(4)}
=\frac{1}{4}(I_{SE}
 +a\sum_{i=1}^{3} \sigma^{(i)}_{S}\otimes \sigma^{(i)}_{E}).\qquad\qquad\qquad\quad
\end{aligned}
\end{equation}
So, from Eq. (\ref{eq:13}), the reference state $\omega_{RSE}=id_{R}\otimes \Lambda_{S} (\omega_{RS})$  is
\begin{equation}
\label{eq:32}
\begin{aligned}
\omega_{RSE}=\sum_{l=1}^{3} \frac{1}{16} \vert l_{R}\rangle\langle l_{R}\vert \qquad\qquad\qquad\qquad\qquad\qquad\qquad\\
\otimes \left( I_{SE}+ \alpha^{(l)}\sigma_{S}^{(l)}\otimes I_{E}
 +a\sum_{i=1}^{3} \sigma^{(i)}_{S}\otimes \sigma^{(i)}_{E}\right) \\
 + \frac{1}{16} \vert 4_{R}\rangle\langle 4_{R}\vert\otimes (I_{SE}
 +a\sum_{i=1}^{3} \sigma^{(i)}_{S}\otimes \sigma^{(i)}_{E}).\qquad\quad
\end{aligned}
\end{equation}

Third, we will show that the $\omega_{RSE}$, in the above equation, is not a Markov state, as Eq.  (\ref{eq:33}). 
For our case, where $S$ is a qubit, there are only three possibilities for decomposing  $\cH_{S}$: $\cH_{S}=\cH_{s^{l}}$,   $\cH_{S}=\cH_{s^{r}}$, and  $\cH_{S}=\cH_{s_{1}} \oplus\cH_{s_{2}}$, where 
$\cH_{s_{1}}$ and $\cH_{s_{2}}$ are one dimensional.
Therefore, a tripartite state $\rho_{RSE}$ is a Markov state if it can be written as $\rho_{RS}\otimes \rho_{E}$, where $\rho_{RS}=\mathrm{Tr_{E}}(\rho_{RSE})$ and $\rho_{E}=\mathrm{Tr_{RS}}(\rho_{RSE})$, or as $\rho_{R}\otimes \rho_{SE}$, where $\rho_{R}=\mathrm{Tr_{SE}}(\rho_{RSE})$ and $\rho_{SE}=\mathrm{Tr_{R}}(\rho_{RSE})$, or as
\begin{equation}
\label{eq:34}
\begin{aligned}
\rho_{RSE}= \lambda_{1}  \rho^{(1)}_{R} \otimes \vert 1_{S}\rangle \langle 1_{S} \vert
\otimes\rho^{(1)}_{E} \quad \\
+ \lambda_{2}  \rho^{(2)}_{R} \otimes \vert 2_{S}\rangle \langle 2_{S} \vert
\otimes\rho^{(2)}_{E},
\end{aligned}
\end{equation}
where $\lbrace \lambda_{1}, \lambda_{2}\rbrace$ is a probability distribution,  $\rho^{(k)}_{R}$ are states on $\cH_{R}$, $\rho^{(k)}_{E}$ are states on $\cH_{E}$, and $\lbrace  \vert 1_{S}\rangle,  \vert 2_{S}\rangle \rbrace$ is an orthonormal basis for  $\cH_{S}$.

Now, from Eq. (\ref{eq:32}), we can verify simply that, for $a\neq 0$, $\omega_{RSE}$ can not be written as 
$\omega_{RS}\otimes \omega_{E}$ or $\omega_{R}\otimes \omega_{SE}$. (For $a=0$, from Eqs. (\ref{eq:30}) and (\ref{eq:32}), we see that $\omega_{RSE}=\omega_{RS}\otimes \frac{1}{2}I_{E}=\omega_{RS}\otimes\omega_{E}$, i.e., $\omega_{RSE}$ is a Markov state.)

 In addition,  we cannot write $\omega_{RSE}$ as Eq. (\ref{eq:34}). For a $\rho_{RSE}$, which can be written as Eq. (\ref{eq:34}), we have  
\begin{equation}
\label{eq:35}
\begin{aligned}
\rho_{RS}=\lambda_{1}  \rho^{(1)}_{R} \otimes \vert 1_{S}\rangle \langle 1_{S} \vert
+ \lambda_{2}  \rho^{(2)}_{R} \otimes \vert 2_{S}\rangle \langle 2_{S} \vert.
\end{aligned}
\end{equation}
From Eq. (\ref{eq:30}), we see that $\langle l_{R} \vert \omega_{RS} \vert l_{R}\rangle=\frac{1}{4} \rho_{S}^{(l)}$. On the other hand, if $\omega_{RS}$ can be written as Eq. (\ref{eq:35}), we have
\begin{equation*}
\label{eq:36}
\begin{aligned}
\langle l_{R} \vert \omega_{RS} \vert l_{R}\rangle=  q_{1} \vert 1_{S}\rangle \langle 1_{S} \vert
+   q_{2} \vert 2_{S}\rangle \langle 2_{S} \vert,
\end{aligned}
\end{equation*}
where $q_{i}=\lambda_{i}\langle l_{R} \vert \rho^{(i)}_{R} \vert l_{R}\rangle$. So, all $\rho^{(l)}_{S}$ must commute with each other. But, from Eq. (\ref{eq:29}), we see that this is not the case. Therefore,  $\omega_{RSE}$ cannot be written as Eq. (\ref{eq:34}). Finally, we conclude that, for $a\neq 0$,  
 the reference state $\omega_{RSE}$, in Eq. (\ref{eq:32}), is not a Markov state, as Eq. (\ref{eq:33}).

Theorem 5 states that the non-Markovianity of $\omega_{RSE}$ leads to the non-CP-ness of the reduced dynamics, for, at least, one $U$. This is in agreement with the result of Refs. \cite{3, 5}. In Ref. \cite{5}, a class of unitary operators as
\begin{equation*}
\label{eq:37}
U=\left( 
\begin{matrix} 
1&0&0&0\\
0 & \mathrm{cos}\, \theta & \mathrm{sin}\, \theta & 0 \\
0 & - \mathrm{sin}\, \theta & \mathrm{cos}\, \theta & 0 \\
0 & 0 & 0 & 1
\end{matrix}
\right) 
%\end{aligned}
\end{equation*}
 has been introduced, where, for some values of $\theta$, the reduced dynamics of the system is non-CP \cite{3, 5}. Note that, even if one shows that the non-CP-ness of the reduced dynamics, for
  %(some values of $\theta$ in)
   the above $U$, is due to inappropriate choosing  the assignment map $\Lambda_{S}$ as  Eq. (\ref{eq:23}), Theorem 5 assures that there exists, at least, one other $U$, for which the reduced dynamics is non-CP, with any possible assignment map $\Lambda_{S}$ (with any possible reference state $\omega_{RSE}$).

It is also worth noting that the above example shows that, even when the reference state $\omega_{RSE}$ is not a Markov state, the reduced dynamics can be CP for some (but not all) $U$, in our case, at least, all $U$ which commute with $\sum \sigma^{(i)}_{S}\otimes \sigma^{(i)}_{E}$.

\section{Summary }\label{sec:summary}

 An straightforward way to construct a convex set of initial states of the system-environment $\mathcal{S}=\lbrace \rho_{SE} \rbrace$
 %, which is the starting point of the framework introduced in Ref. \cite{3},
  is to consider the set of steered states, from a reference state   $\omega_{RSE}$.
In Sec. ~\ref{sec:steered set}, we have shown that if $\omega_{RSE}$ can be written as Eq. (\ref{eq:2a}), then the reduced dynamics of the system is Hermitian. For the special case that the assignment map $\Lambda_{S}$, in Eq.  (\ref{eq:2a}), is CP, the reduced dynamics is so CP. Interestingly, this includes all the previous results in this context, in Refs. \cite{11, 12, 14, 13, 15, 16}.

The convex set of initial states  $\mathcal{S}=\lbrace \rho_{SE} \rbrace$ is the starting point of the framework introduced in Ref. \cite{3}. From this $\mathcal{S}$, we can construct the subspace $\mathcal{V}\subseteq \mathcal{L}(\cH_{S}\otimes\cH_{E})$.
Now, in Sec. ~\ref{sec:ref state}  (Sec. ~\ref{sec:generalization}), we have shown that $\mathcal{V}$  ($\mathcal{V}^{\prime}$) and $\mathcal{V}_{S}=\mathrm{Tr_{E}}\mathcal{V}$
 can be written as the generalized steered sets, from the reference states  $\omega_{RSE}$ and $\omega_{RS}$, in Eqs. (\ref{eq:12}) and  (\ref{eq:13}), respectively. The relation between $\omega_{RSE}$ and $\omega_{RS}$ is as Eq.  (\ref{eq:2a}). Therefore, the steered set, from a reference state as Eq.  (\ref{eq:2a}), gives us the most general set  (within the framework of Ref. \cite{3}) for which the reduced dynamics is Hermitian.
 
In addition, the evolution of the system-environment (system) states is given by the evolution of the reference state, in Eq. (\ref{eq:15}) (Eq. (\ref{eq:17})).  Interestingly, for a unitary evolution of the system-environment $U$, the reduced dynamics of the system is CP if and only if $\omega_{RS}(t)$ can be written as Eq. (\ref{eq:12b}), with a CP map $\mathcal{E}_{S}^{(CP)}(t)$.

%A general framework for  Hermitian maps, arisen from  Eq. (\ref{eq:two}), has been developed in Ref. \cite{3}. The starting point of this framework is to consider a convex set of initial states $\mathcal{S}=\lbrace \rho_{SE} \rbrace$, for the whole system-environment. There, it is not determined how to construct such convex $\mathcal{S}$. An straightforward way to do so is to consider the set of steered states, from a  reference state   $\omega_{RSE}$.

This fact that we can construct reference state  $\omega_{RSE}$, for arbitrary $U$-consistent subspace,
 %(which gives us the most general set of initial system-environment states, within the framework of Ref. \cite{3}),
  leads us to an important result, i.e., the generalization of the result of Ref. \cite{15}, to arbitrary $\mathcal{U}(\cH_{S}\otimes\cH_{E})$-consistent $\mathcal{V}$: The reduced dynamics of the system, for arbitrary system-environment unitary evolution $U$, is CP if and only if the reference state $\omega_{RSE}$, in Eq. (\ref{eq:13}), is a Markov state, as Eq. (\ref{eq:33}).

Finally, in Sec. ~\ref{sec:example}, we have considered the case studied in Ref. \cite{5}.
% to illustrate (a part of) our results. 
This example illustrates this result that when the reference state $\omega_{RSE}$ is not a Markov state, then the reduced dynamics is non-CP, for at least one $U$. In addition, this example shows that, even when 
$\omega_{RSE}$ is not a Markov state, the CP-ness of the reduced dynamics, for some (but not all) $U$, is possible.

% two immediate consequences of introducing : If $\omega_{RSE}$, in Eq.  (\ref{eq:13}),   is not a , then
%it is impossible to find a completely positive assignment map, and, in addition,
% the reduced dynamics of the system, for at least one $U$, is not completely positive.

%%%%%%%%%%%%%%%%%%%%%%%%%%

\end{document}